\documentclass[prb,twocolumn,showpacs,superscriptaddress]{revtex4}

\usepackage{epsfig}
\begin{document}

\title{Silicene Beyond Mono-layers - Different Stacking Configurations And Their Properties}
\author{C. Kamal\footnote{Corresponding author: C. Kamal; e-mail:ckamal@rrcat.gov.in }}
\affiliation{ Indus Synchrotrons Utilization Division, Raja Ramanna Centre for Advanced Technology, Indore 452013, India }
\author{Aparna Chakrabarti}
\affiliation{ Indus Synchrotrons Utilization Division, Raja Ramanna Centre for Advanced Technology, Indore 452013, India }
\author{Arup Banerjee}
\affiliation{BARC Training School, Raja Ramanna Centre for Advanced Technology, Indore 452013, India }
\author{S. K. Deb}
\affiliation{ Indus Synchrotrons Utilization Division, Raja Ramanna Centre for Advanced Technology, Indore 452013, India }
  
\begin{abstract}
We carry out a computational study on the geometric and electronic properties of multi-layers of silicene in different stacking configurations using a state-of-art \textit{abinitio} density functional theory based calculations. In this work we investigate the evolution of these properties with increasing number of layers ($n$) ranging from 1 to 10. Though, mono-layer of silicene possesses properties similar to those of graphene, our results show that the geometric and electronic properties of multi-layers of silicene are strikingly different from those of multi-layers of graphene. We observe that there exist strong inter-layer covalent bondings between the layers in multi-layers of silicene as opposed to weak van der Waal's bonding which exists between the graphene layers. The inter-layer bonding strongly influences the geometric and electronic structures of these multi-layers. Like bi-layers of graphene,  silicene with two different stacking configurations AA and AB exhibits linear and parabolic dispersions around the Fermi level, respectively. However, unlike graphene, for bi-layers of silicene, these dispersion curves are shifted in band diagram; this is due to the strong inter-layer bonding present in the latter. For $n >$ 3, we study the geometric and electronic properties of multi-layers with four different stacking configurations namely, AAAA, AABB, ABAB and ABC. Our results on cohesive energy show that all the multi-layers considered are energetically stable. Furthermore, we find that the three stacking configurations (AAAA, AABB and ABC) containing tetrahedral coordination have much higher cohesive energy than that of Bernal (ABAB) stacking configuration. This is in contrast to the case of  multi-layers of graphene where ABAB is reported to be the lowest energy configuration.  We also observe that bands near the Fermi level in lower energy stacking configurations AAAA, AABB and ABC correspond to the surface atoms and these surface states are responsible for the semi-metallic character of these multi-layers.

\end{abstract}

\maketitle

\section{Introduction}

Finding novel materials is one of the prime goals of material science research activity. With the advance of sophisticated experimental methods and characterization techniques, many novel materials have been synthesized and their properties have also been investigated. Graphene is one such material with many novel properties and it has been extensively studied both theoretically and experimentally\cite{}. The charge carriers in this two-dimensional system behave like massless relativistic Dirac Fermions giving rise to a linear dispersion around the Fermi energy at a highly symmetric k-point ($K$) in the reciprocal lattice. This system exhibits various other properties such as anomalous integer quantum Hall effect,  Klein tunneling, and non-zero minimum DC conductivity\cite{grap1,grap3,grap4,grap5}. Recently,  silicon based nanostructures such as silicene ( the silicon counter part of graphene) and silicene nanoribbons have attracted the interest of many researchers\cite{sili-ciraci1,sili-ciraci2,ni,drummond,eza1,ck-arxive,eza2,eza3,sili-soc,sili2,sili3,refpap1,refpap2,refpap3,refpap5,sili-grown, sili-expt,lay,sili-expt4,sili-expt5,sili-expt6,sili-multi,sinr1,sinr2} due to the properties which are similar to those of graphene. 
Moreover, Silicon based nanostructures have some distinct advantages over carbon based nanostructures. The former systems are expected to be compatible with the existing semiconductor industry. Furthermore, it has been theoretically predicted that a band gap can be opened up and tuned in mono-layer of silicene by applying an external transverse electric field\cite{ni,drummond,eza1,ck-arxive}. On the other hand, inducing a band gap by applying an electric field is not possible in mono-layer of graphene. It is interesting to note that, recently, silicene has been grown epitaxially on a close-packed silver surface \cite{sili-grown, sili-expt,lay,sili-expt4,sili-expt5,sili-expt6,sili-multi} and hence it opens up a possibility of validating the existing theoretical prediction.

It is observed that the electronic structure of silicene possesses linear dispersion around Dirac point which is similar to that of graphene and hence it is a potential candidate for applications in nanotechnology\cite{sili-ciraci1,sili-ciraci2,ck-arxive,ni,drummond,eza1,eza2,eza3,sili-soc,sili2,sili3,refpap1,refpap2,refpap3,refpap5}. It is important to note that, the geometric  structure of mono-layer of silicene is slightly different from the planar structure of mono-layer of graphene. The structure of mono-layer of silicene is buckled and the presence of this buckling results in increase in the cohesive energy of system. The reason for buckling is due to the mixing of sp$^2$ and sp$^3$ hybridizations rather than purely sp$^2$ hybridization. This is due to fact that silicon favors sp$^3$ hybridization.  

In the literature, there have been many theoretical and experimental studies on bi- and multi-layers of graphene in the recent past\cite{grap-bilayers1,grap-bilayers2,grap-bilayers3,grap-bilayers4,grap-bilayers6,grap-bilayers7,grap-bilayers-disp,grap-multilayers1,grap-multilayers2,grap-multilayers3,grap-multilayers4}. 
The studies on bi-layer of graphene indicate that it possesses parabolic dispersion around the highly symmetric k-point ($K$) in the reciprocal lattice as opposed to the linear dispersion in the case of mono-layer. Though mono-layer of graphene possesses many interesting properties, the bi-layer structure is much more important from the application point of view. For example, there is no band gap in the pristine  mono- and bi-layer of graphene. However, it has been shown both theoretically and experimentally that a  band gap can be opened up in bi-layer of graphene by applying a gate voltage. Further, the value of band gap can be tuned over a wide range which may have potential applications in nanoelectronics and nanodevices\cite{grap-bilayers1,grap-bilayers2}. 
It is to be  noted  that there are numerous studies on bi- and multi-layers of graphene, however, detailed theoretical studies on similar systems of silicene are lacking in the literature. Hence, it can be interesting to study the properties of bi- and multi-layers of silicene, given the importance of silicon 
based nanostructures.  In this work, we carry out a detailed investigation of geometric and electronic properties of bi- and multi-layers of silicene with different stacking configurations using state-of-art \textit{abinitio} density functional theory based calculations. We also study the evolution of geometric as well as electronic structures of multi-layers of silicene with increasing number of layers ($n$ = 1  to 10).  
In the next section, we briefly outline the computational methods employed in the present work. The results and discussions are presented in section III and then followed by conclusion in section IV.

\section{Computational details}

We perform density functional theory (DFT)\cite{dft} based calculation using Vienna ab-initio simulation package (VASP)\cite{vasp} within the framework of the projector augmented wave (PAW) method. We use generalized gradient approximation (GGA) given by Perdew-Burke-Ernzerhof (PBE)\cite{pbe} for exchange-correlation potential. The cutoff for the plane wave expansion is taken to be 400 eV and the mesh of k-points for Brillouin zone integrations is chosen to be 21$\times$21$\times$1. The convergence for plane wave cutoff and number of k-points in the mesh have been checked by varying these parameters. The convergence criteria for energy in SCF cycles is chosen to be 10$^{-6}$ eV. All the structures are optimized by minimizing the forces on individual atoms with the criterion that the total force on each atom is below 10$^{-2}$ eV/ $\AA$. We use a super-cell geometry with a vacuum of about 15  $\AA$  in the z-direction (direction perpendicular to the plane  of silicene) so that the interaction between two adjacent unit cells in the periodic arrangement is negligible. All the geometric structures and charge density distributions are plotted using XCrySDen software\cite{xcrysden}.

\section{Results and Discussions}
\subsection{Geometric Structure and Cohesive Energy} 
\begin{figure}[]
\begin{center}
\includegraphics[width=7cm]{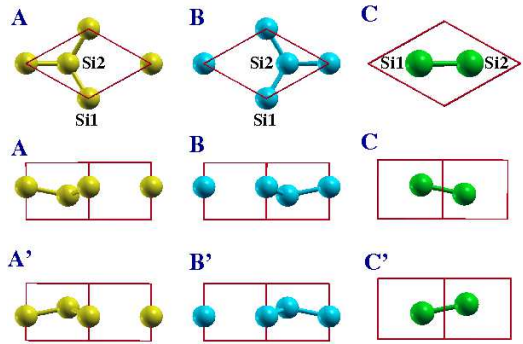}
 \caption{(Color online) Three basic building blocks of mono-layer of silicene in which two Si atoms are at highly symmetric positions. Top view (top panel) and side view (middle and bottom panels) }
\end{center}
 \end{figure}

The unit cell of mono-layer of silicene has two Si atoms (Si1 and Si2) which forms a hexagonal lattice with space group of P3m1. In this work, we consider three possible arrangements for these two Si atoms which are placed at symmetric positions. The geometric structures of these three basic building blocks are shown in Fig. 1. These arrangements are (i) A: Si1 (0, 0, 0), Si2 (2/3, 1/3, z), (ii) B: Si1 (0, 0, 0), Si2 (1/3, 2/3, z), and (iii) C: Si1 (2/3, 1/3, 0), Si2 (1/3, 2/3, z). The variable 'z' in the fractional coordinates along the z-axis indicates that the two silicon atoms in the unit cell are not in the same plane and this is due to the effect of buckling. The relative position of Si2 can be below (A, B and C) or above (A', B' and C') the Si1 atom. The multi-layers of silicene with different stacking configurations are constructed with these three building blocks. In the present work, we consider following four different stacking configurations : (1) AAAA - simple hexagonal, (2) AABB - double hexagonal, (3) ABC - rhombohedral, and (4) ABAB - Bernal  stacking. The first three stacking configurations lead to a tetrahedral arrangement of Si atoms which is a favourable configuration for silicon based systems. In case of multi-layers of graphene, the Bernal stacking corresponds to the minimum energy configuration and hence we include this stacking as well to study the multi-layers of silicene. We carry out the geometry optimization of multi-layers of these four stackings with number of layers up to ten (from mono-layer ($n$=1) to ten layers ($n$=10)). The optimized geometries of ten layers with all the four stacking configurations are shown in Fig. 2. 

\begin{figure}[]
\begin{center}
\includegraphics[width=7cm]{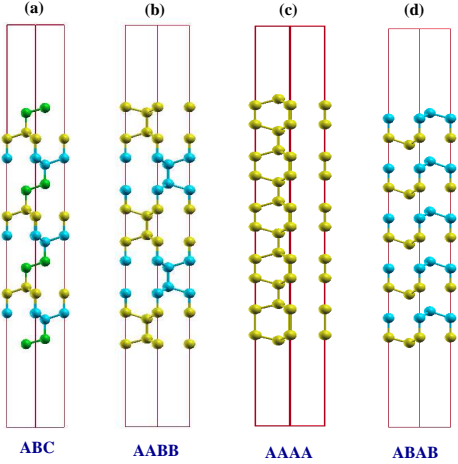}
 \caption{(Color online) The optimized geometric structures of multi-layers  ( $n$ = 10 ) with four stacking configurations : (a) ABC, (b) AABB, (c) AAAA and (d) ABAB. }
\end{center}
 \end{figure}

\textit{Cohesive Energy:} 
In order to study the stability of the multi-layers of silicene with different stackings, the cohesive energy $E_c$ of all the multi-layers has been calculated by using the expression 
\begin{equation}
E_c = 2 n E_{Si} - E_n,
\end{equation}
where $ E_n$ and $E_{Si}$ are the energies of multi-layers ( $n$- layers) and Si atom respectively. In the above expression, $E_{Si}$ is multiplied by 2 since there are two Si atoms in each layer of the unitcell. The variation in values of cohesive energy per atom with increasing number of layers for different stackings is plotted in Fig. 3. We observe that all the multi-layered structures considered in this work are energetically stable. Our calculations suggest that the rhombohedral stacking (ABC) is the minimum energy configuration. However, the values of cohesive energy of ABC stacking are much closer to those of AAAA, and AABB stackings. The cohesive energy per atom of 10 layers in AABB and AAAA stackings are 7 and 15 meV/atom lower than that of ABC stacking, respectively. 
The cohesive energies per atom of three stacking configurations, AAAA, AABB and ABC increase smoothly with the number of layers. Interestingly, cohesive energy of Bernal stacking (ABAB) is much lower than those of all the other stackings. In case of 10 layers, the cohesive energy per atom of ABAB stacking configuration is 236 meV/atom lower than that of ABC stacking. This result is in contrast to that of multi-layers of graphene where ABAB stacking is the lowest energy configuration. We also observe an oscillation in the values of cohesive energy of Bernal stacking. In order to understand these results for cohesive energies of multi-layers of silicene, we carry out a detailed investigation on the geometric structures of all the multi-layers of silicene.
\begin{figure}[]
\begin{center}
\includegraphics[width=7cm]{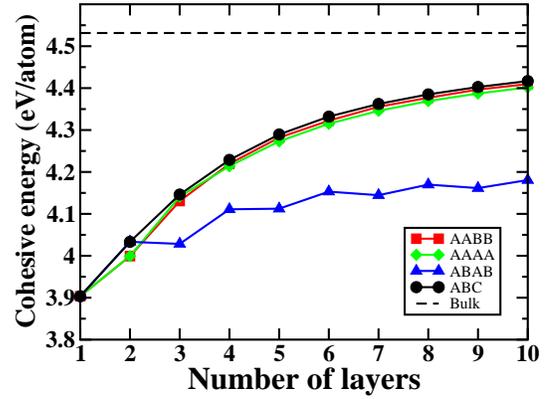}
 \caption{(Color online) Variation of the cohesive energy per atom with the increasing number of layers for the four stackings ABAB (with triangle symbol), AAAA (with diamond symbol),  AABB (with square symbol) and ABC (with circle symbol). The dash line indicates the value of cohesive energy of hexagonal bulk Silicon.}
\end{center}
 \end{figure}

\subsubsection{Mono-layer}
 In mono-layer of silicene, the two Si atoms have three nearest neighbors with bond length of 2.279 $\AA$ and hence these atoms make three strong $\sigma$ bonds and one weak $\pi$ bond with their nearest-neighbors. Furthermore, the $\pi$ bonds in silicene are weak due to large internuclear distance between Si atoms (d$_{Si-Si}$ is much larger than d$_{C-C}$ =1.42 $\AA$ in graphene). In Table. I, we compile the results for the lattice constant, bond length and bond angle corresponding to the optimized structures of mono- and bi-layer of silicene. We note that these results agree well with those available in the literature\cite{sili-ciraci1,sili-ciraci2}. It is important to note that the hybridizations in graphene (sp$^2$) and silicene (mixture of sp$^2$ + sp$^3$) are different due to the presence of buckling. 

\subsubsection{Bi-layers}
 In case of bi-layers of silicene, our results on geometric structures show that there exist a strong coupling between the layers of silicene as opposed to a weak van der Waal's interaction which exists in the multi-layers of graphene and graphite. For bi-layers, there are only two possible stackings namely, AA and AB. We perform geometry optimization of the bi-layers of silicene with starting inter-layer separation equivalent to that of bi-layers of graphene. The optimized structures of these bi-layers are displayed in Fig. 4.  These results show that the two layers
are covalently bonded with each other. Consequently, the $E_c$ per atom for bi-layers of silicene increases significantly from that of mono-layer. It is also observed that AB stacking is lower in energy as compared to the AA stacking and difference in cohesive energy per atom between these two structures is 34 meV/atom. 
In order to further examine the nature of bonding between the two layers of silicene, we study the valence charge density distributions of bi-layers. In Fig. 5, we show the valence charge density distributions of AA and AB stacked bi-layers of graphene ((a) and (c)) and silicene ((c) and (d)). The analysis of valence charge density distribution in both AA and AB stackings of  silicene corroborates to the results of optimized geometries discussed above. The significant charge density distribution around the inter-layer bonds establishes the covalent nature of these bonds. This is different from a weak van der Waal's coupling which exists between two graphene layers (see Fig. 5(a) and (b)) 
We also observe that the buckling and the bond lengths in both AA and AB stacked bi-layers have increased compared to corresponding results for the mono-layer of silicene. Furthermore, the results presented in Table. I clearly show that there is a reduction in the values of bond angles between atoms in same layer (intra-layer) and there is an increase in the values of bond angles between atoms in two layers (inter-layer). These results suggest that the contribution of sp$^3$ hybridization has increased as compared to that of mono-layer of silicene.

\begin{figure}[]
\begin{center}
\includegraphics[width=7cm]{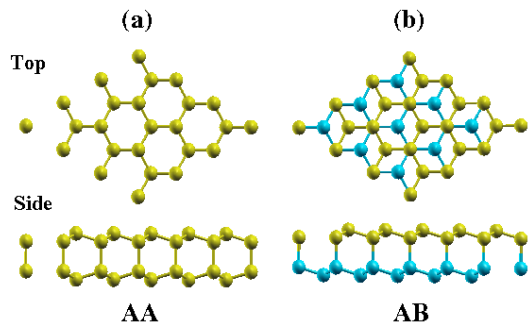}
 \caption{(Color online)  The optimized geometric structures of (a) AA and (b) AB stacked bi-layers of silicene. Top and side view of 3 $\times$3 super cell.}
\end{center}
 \end{figure}

\begin{figure}[]
\begin{center}
\includegraphics[width=7cm]{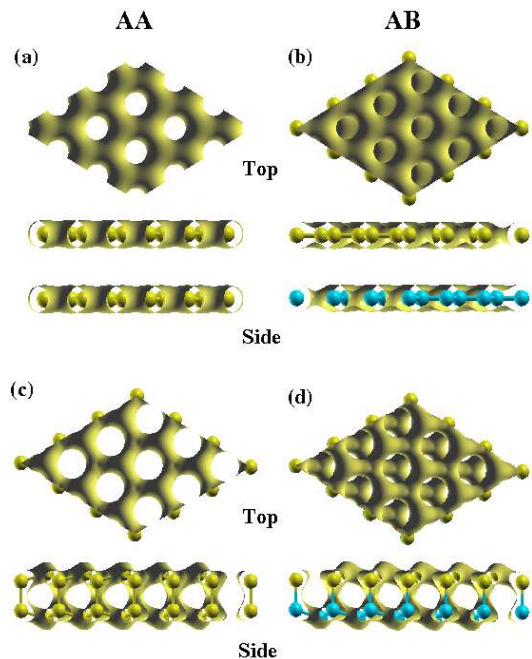}
 \caption{(Color online) The valence charge density distributions of AA and AB stacked bi-layers of graphene ((a) and (b)) and silicene ((c) and (d)). Top and side view of 3 $\times$3 super cell.}
\end{center}
 \end{figure}

\begin{table}
 \caption{The results of optimized geometries of mono- and bi-layers of silicene obtained by DFT with PBE exchange-correlations functional.}
\begin{scriptsize}
\begin{tabular}{lcccc}
 \hline
\hline
System	&	Lattice 	&	Bond  &	Bond 	&	Buckling  \\
	      &	Constant	&	Lengths &	Angles	&	Length  \\
                 &       ($\AA$)  &       ($\AA$)        & ($^\circ$)  & ($\AA$) \\
\hline
\hline
Mono-layer	&	3.867	&	2.279		&	116.08	&	0.457	\\
Bi-layer AB	&	3.851	&	2.321, 2.528	&	112.10, 106.69	&	0.667	\\
Bi-layer AA	&	3.858	&	2.324, 2.464	&	112.21, 106.57	&	0.663	\\
\hline
\end{tabular}
\end{scriptsize}
\end{table}

\subsubsection{Multi-layers}
 Now, we discuss the results for optimized geometry and cohesive energy of multi-layers of silicene with four different stacking configurations.  In Fig. 2, we show the optimized structures of multi-layers ($n = $ 10). This figure clearly elucidates the existence of a inter-layer covalent bonding in the multi-layers of silicene similar to the case of bi-layers.  As discussed above, the cohesive energies of multi-layers with the three stacking configurations AAAA, AABB and ABC are significantly higher than that of the structure with ABAB stacking. This can be attributed to the fact that all the inner layers are covalently bonded with the adjacent layers in these three stackings. Due to this inter-layer bonding, all the silicon atoms in these three stackings, except those on surfaces, have four nearest neighbors ( three intra-layer and one inter-layer) in nearly tetrahedral configurations. Hence, all the valence electrons in the silicon atoms (except ones on surface) make four sigma bonds with their nearest neighbors, which leads to  sp$^3$-like hybridization in these stackings.  On the other hand, in ABAB stacking, Si1 atom in layer A makes a fourth inter-layer sigma bond with Si1 atom in layer B along z-direction (see Fig. 2, both the atoms are at same (x, y) ) and hence these atoms assume nearly tetrahedral configuration. However, Si2 atoms in layers A and B do not have strong sigma bonds with each other ( they are at different (x, y)) and Si2-Si2 distance is more than 3 $\AA$.  Each set of AB layers are well connected by Si1-Si1 bonds but there are two Si2 atoms which lack the  fourth nearest neighbors to form the sigma bond. Therefore, the number of sigma bonds in this stacking is less and hence they have lowest cohesive energy as compared to those of AAAA, AABB and ABC stacking configurations.  The reason for closeness of cohesive energies of  AAAA, AABB and ABC stacking configurations is the similar bonding environment of each Si atom in these three stackings.   

 \textit{Valence Charge Density: } Having discussed the geometric properties of multi-layers of silicene, we now present the results for the valence charge density distribution of these multi-layers.  In Fig. 6, we plot the valence charge density distribution for six layers with ABC and ABAB stacking configurations. 
This figure clearly brings out the differences in bondings properties in these two configurations, which are consistent with our results obtained from the optimized geometries of these multi-layers.  Furthermore, we note here that the cohesive energy of ABAB stacking configuration show oscillations with increasing number of layers. This can be explained by examining the charge density distribution of this stacking configuration as shown in Fig. 6 (b). It is observed that each bi-layer (AB) in ABAB stacking is strongly connected by inter-layer Si1-Si1 covalent bonds. However,  the bonding between two adjacent AB bi-layers is weak and they are connected by the inter-layer Si2 atoms. The valence charge density presented in Fig. 6(b) confirms the presence of strong Si1-Si1 and weak Si2-Si2 bonds. Hence, the structures with even number of layers have large cohesive energy as compared to the structures with odd number of layers. This leads to an oscillation in cohesive energy per atom in ABAB stacking.  
\begin{figure}[]
\begin{center}
\includegraphics[width=7cm]{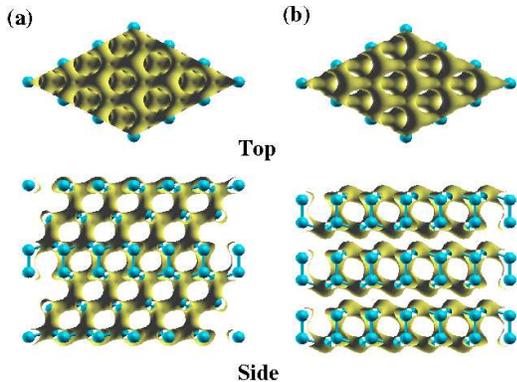}
 \caption{(Color online) The valence charge density distributions of multi-layers of silicene ($n$ = 6) with (a) ABC and (b) ABAB stacking configurations.}
\end{center}
 \end{figure}

\textit{Evolution of Properties: } We also study the evolution of the structural properties of multi-layers with increasing number of layers. To this end, we plot in Fig.7 the variations of (a) lattice constant, (b) thickness  and (c) difference of thickness of multi-layers with AAAA and AABB stackings with respect to that of lower energy ABC stacking as a function of $n$. In this plot, we consider only the lowest energy stacking configurations. It can be clearly seen from Fig. 7(a) that the lattice constant of ABC stacking is always higher than those of AAAA and AABB. Consequently, the geometric structure of multi-layer with the ABC stacking is slightly more relaxed in lateral direction than AAAA and AABB stacking configurations. For a given number of layers the ABC stacked structures are thick
as compared to AAAA and AABB structures. The slightly higher relaxation in lateral and the compression in transverse direction of ABC stacked structures may have led to the small increase in the cohesive energy of this stacking as compared to those of AAAA and AABB stackings.

\begin{figure}[]
\begin{center}
\includegraphics[width=7cm]{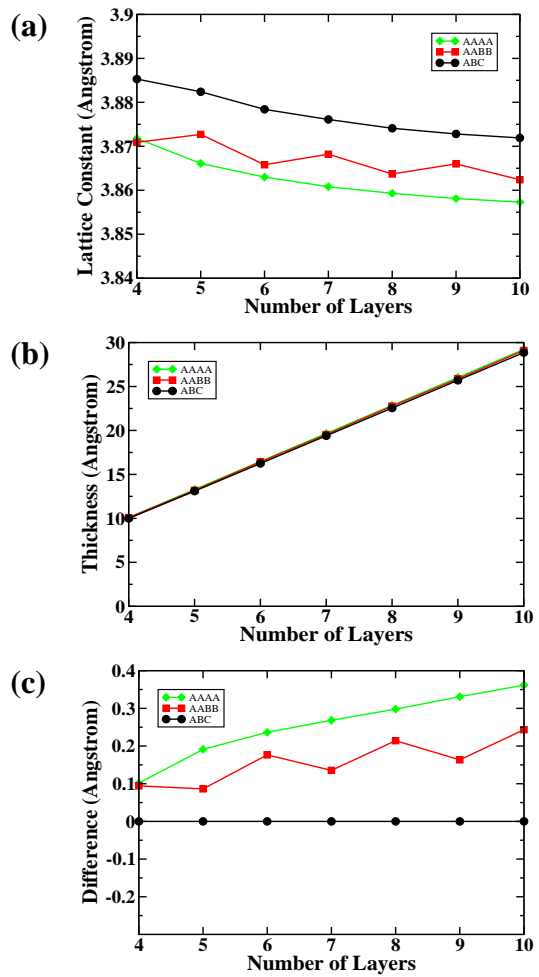}
 \caption{(Color online) Variation of (a) the lattice constant, (b) thickness and (c) difference of thickness of multi-layers with AAAA and AABB stackings with respect to that of ABC stacking with increasing number of layers. }
\end{center}
 \end{figure}

\subsection{Electronic Structures}

Having discussed the results for optimized geometric structures of  multi-layers of silicene  in previous section, we now focus our attention on the electronic properties of these systems. For this purpose we calculate and study the band structure of multi-layers of silicene with number of layers ranging from 1 to 10. 

\subsubsection{Band Structure of Mono-layer}
First, we start our discussion on the results of the band structure which are already established in the literature\cite{sili-ciraci1,sili-ciraci2}. The plot of band structure of mono-layer of silicene along the highly symmetric k-points in Brillouin zone is given in Fig. 8 (a). This result clearly shows linear dispersion around high symmetry k-point ($K$) near the Fermi level E$_F$ (see enlarged dispersion in Fig. 8 (d)). We infer that the mono-layer of silicene is a semi-metal with the valence and conduction bands touching each other at the highly symmetric k-point ($K$). From the analysis of angular momentum (l) projected bands (not shown here), we confirmed that the valence and conduction bands near E$_F$  are mainly due to $\pi$ and $\pi^*$ orbitals respectively. It is also observed that there is a small mixing of $\pi$ and $\sigma$ states (or mixture of sp$^2$and sp$^3$ hybridizations) which leads to a buckling of the mono-layer of silicene. 
Our results on mono-layer match well with those available in the literature\cite{sili-ciraci1,sili-ciraci2}
\begin{figure}[]
\begin{center}
\includegraphics[width=7cm]{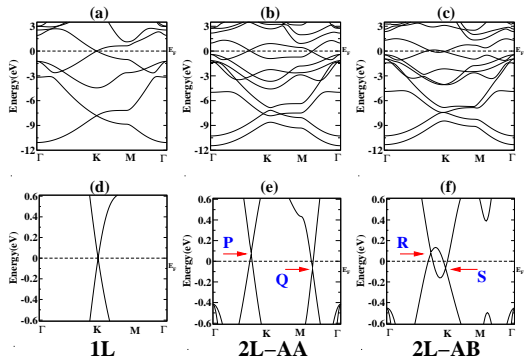}
 \caption{ Band structures of Mono-layer ((a) and (d)) and Bi-layers ((b), (c), (e) and (f)) of silicene in two different energy ranges.}
\end{center}
 \end{figure}

\begin{figure*}
\begin{center}
\includegraphics[width=15cm]{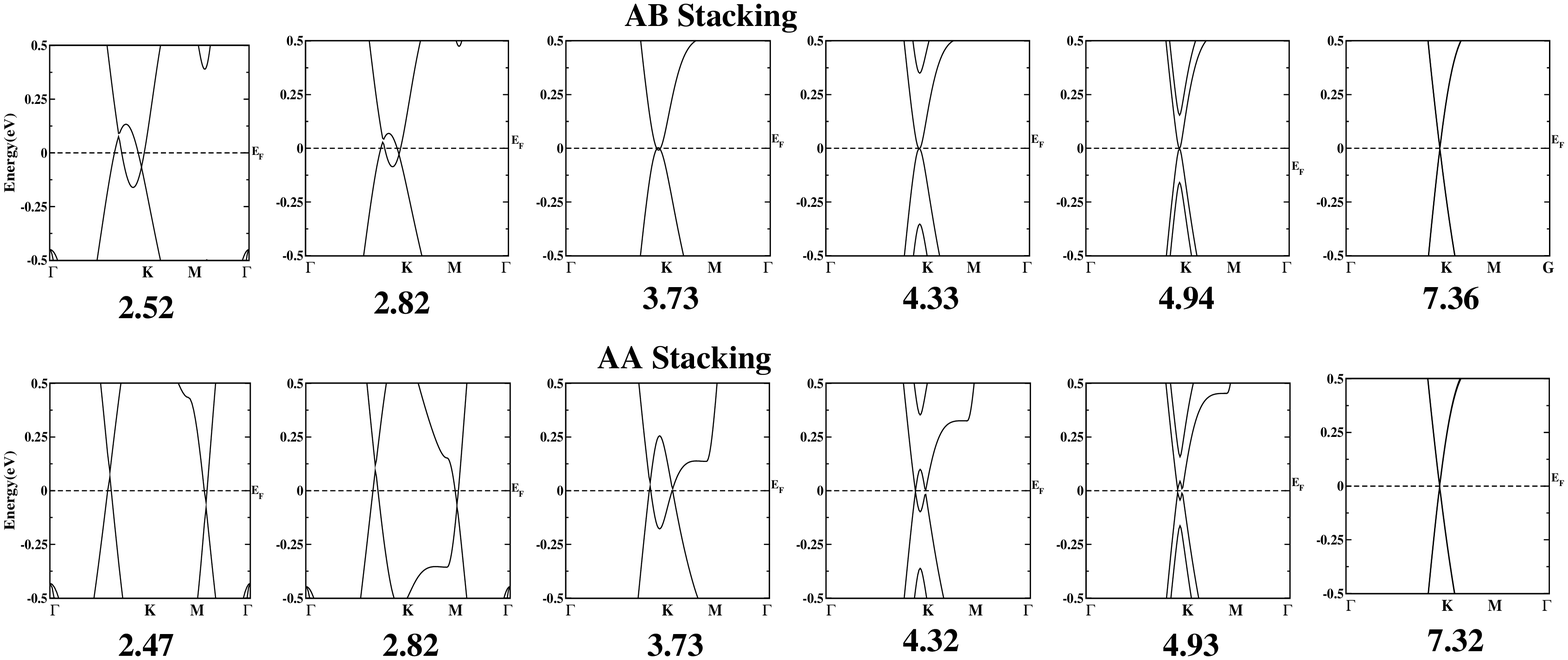}
 \caption{Variation of band structures of AA and AB stacked bi-layers of silicene with different inter-layer separations. A closer look reveals that for an inter-layer separation of about 3.73 $\AA$,  the band structures of silicene become similar to those of bi-layers of graphene.}
\end{center}
 \end{figure*}

\subsubsection{Band Structure of Bi-layers}

As mentioned earlier, there are two possible stacking structures namely AB and AA for bi-layers of silicene. The band structure for AB and AA stackings are shown in Fig. 8 (b) and (c) respectively. It is clearly seen from the figures that dispersions around E$_F$ for these two cases are different and hence the electronic properties strongly depend upon the nature of stackings. For AA configuration, it is observed that there exists linear dispersion along two directions namely, $\Gamma$-K and $\Gamma$-M ( see Fig. 8 (e)). Each of these dispersion curves crosses at two points, denoted by P and Q in the band diagram (E-k). On the other hand for AB stacked bi-layer, we obtain a  dispersion which is parabolic in nature ( see Fig. 8 (f)). These two dispersion curves also cross each other at two points namely R and S in the E-k diagram. It is interesting to note that all these four points do not coincide with the Fermi level. In particular, the points P and R lie above E$_F$ while the points Q and S lie below E$_F$. The presence of linear and parabolic dispersions in AA and AB stacked bi-layers respectively is similar to those present in bi-layers of graphene\cite{}. However, in case of bi-layers of graphene, all these crossing points (P, Q, R and S) lie at E$_F$. The differences in the band structures of bi-layers of silicene from that of graphene may be due to the presence of strong inter-layer covalent bonding  in bi-layers of silicene as compared to the weak van der Waal's bonding between the layers of graphene. In order to verify this, we carry out calculations of band structure for both AA and AB stacked bi-layers of silicene with different inter-layer separations and these results are presented in Fig. 9.

\textit{Variation of Band Structure with Inter-layer Separation: }
  It can be clearly observed from Fig. 9 that these four crossing points (P, Q, R and S) slowly move toward each other in E-k diagram (both E and k undergo change) and finally merge with each other at E$_F$ for a distance larger than 7 $\AA$ between the two layers. Furthermore, the band structures of both AA and AB stacked bi-layers are reduced to that of mono-layer since there is no interaction between the layers at a distance larger than 7 $\AA$.  We also observe during the evolution of band structure of bi-layers of silicene with increasing inter-layer separation that the band structures become similar to those of bi-layers of graphene for the intermediate inter-layer distances of about 3-4 $\AA$. Therefore, we note that the origin of the differences in the band structures of bi-layers of silicene and graphene is indeed due to the presence of strong inter-layer coupling in the former. 
 To estimate the strength of coupling, we calculate the splitting of energies of top most valence bands at highly symmetric k-point $\Gamma$. Note that the first and second valence bands are degenerate and similar is the case for the third and fourth bands at $\Gamma$ point. Therefore, we estimate the amount of splitting by calculating the difference in the energies between the first and third bands at $\Gamma$ point. The values of splitting for AA and AB stacked bi-layers are found to be 0.748 and 0.502 eV respectively . These values indicate that the splitting is more in AA stacking and hence experiences much stronger inter-layer interaction as compared to that in AB stacking. These results are consistent with our results on the geometric structure that the inter-layer bond length in AA stacking (2.464 $\AA$	) is shorter than that of AB stacking (2.538 $\AA$).

\subsubsection{Band Structure of Multi-layers}

\begin{figure}[]
\begin{center}
\includegraphics[width=5cm]{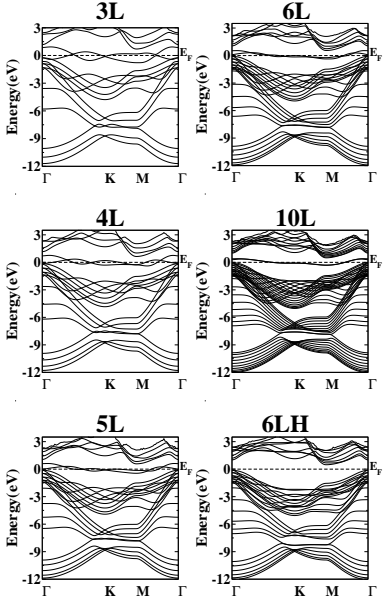}
 \caption{Evolution of band structures of AAAA stacked  multi-layers of silicene with increasing number of layers.}
\end{center}
 \end{figure}

\begin{figure}[]
\begin{center}
\includegraphics[width=5cm]{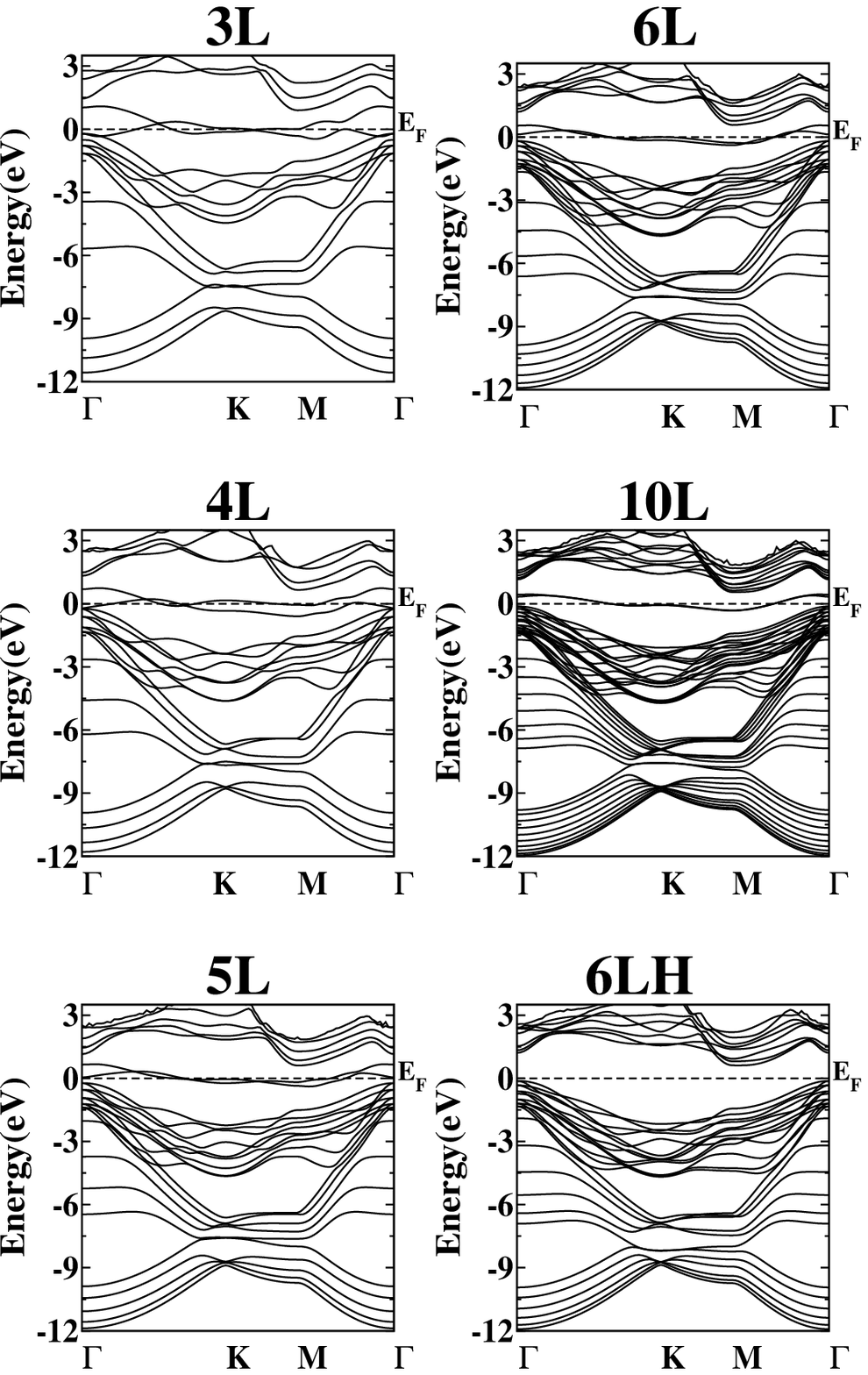}
 \caption{Evolution of band structures of AABB stacked multi-layers of silicene with increasing number of layers.}
\end{center}
 \end{figure}

\begin{figure}[]
\begin{center}
\includegraphics[width=5cm]{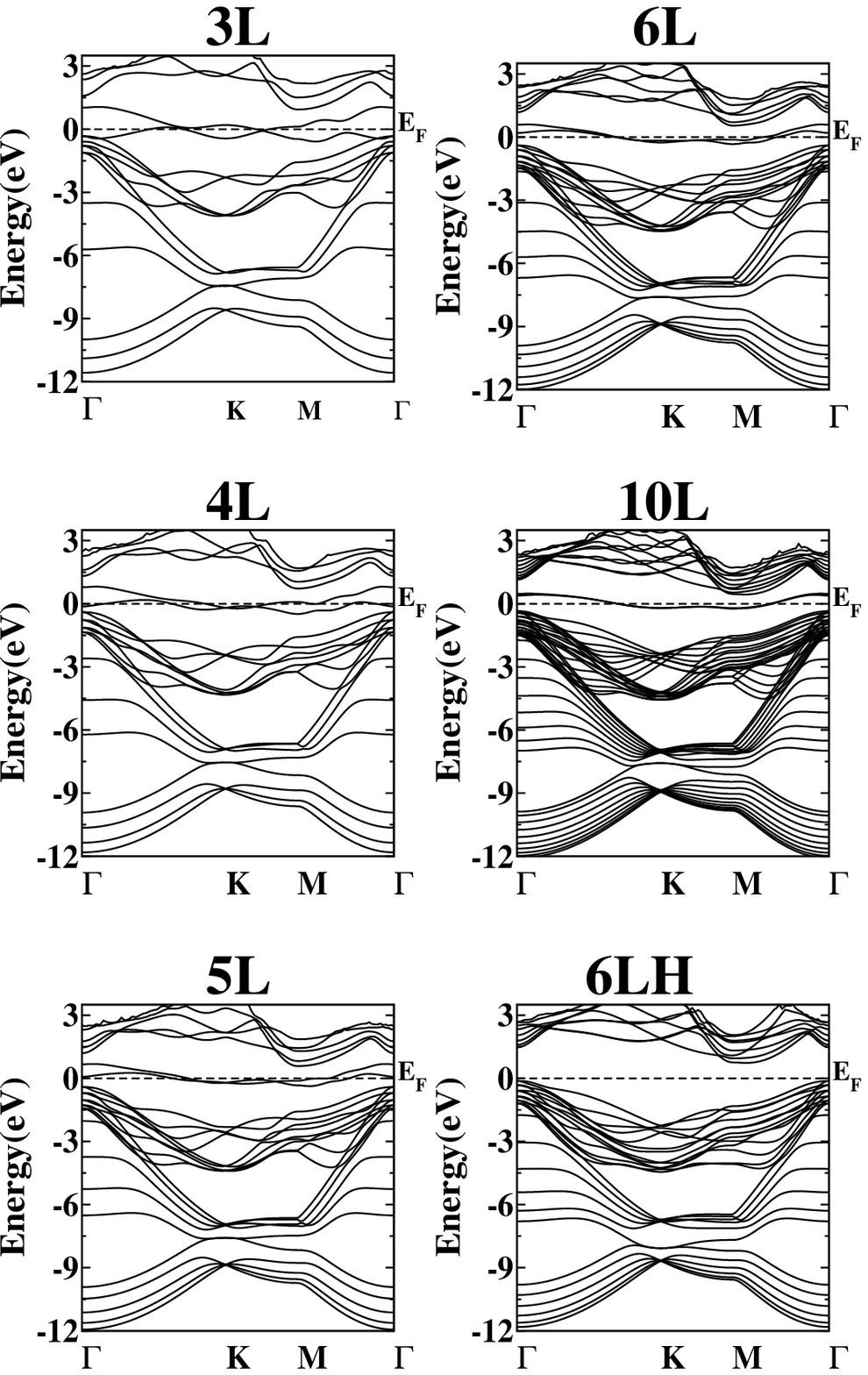}
 \caption{Evolution of band structures of ABC stacked multi-layers of silicene with increasing number of layers.}
\end{center}
 \end{figure}

\begin{figure}[]
\begin{center}
\includegraphics[width=5cm]{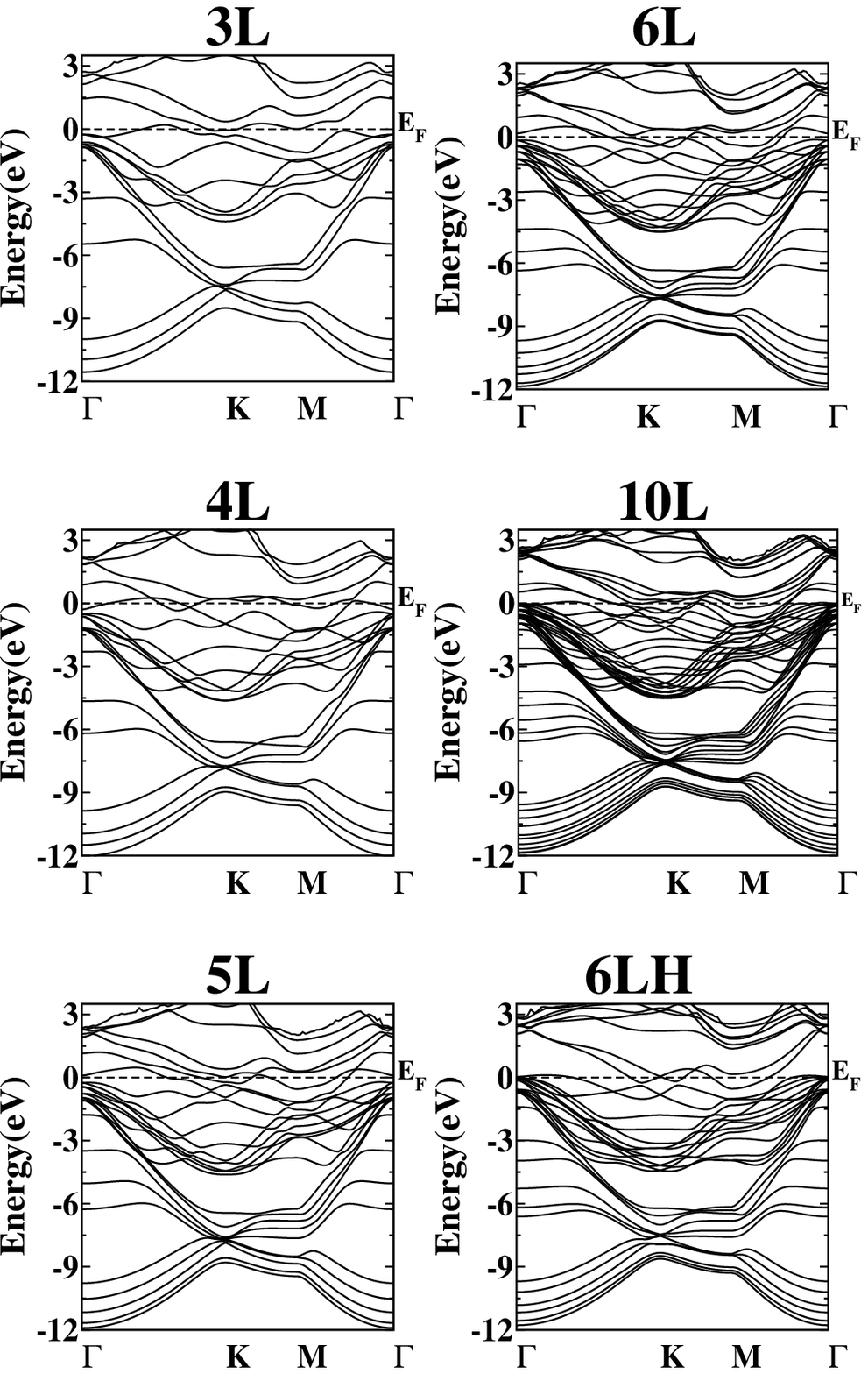}
 \caption{Evolution of band structures of ABAB stacked multi-layers of silicene with increasing number of layers.}
\end{center}
 \end{figure}

The interesting trends found in the bi-layer prompt us to probe the band structures of multi-layers of silicene with more than two layers in four different stacking configurations namely, AAAA, AABB, ABC and ABAB.  For each stacking configuration, we plot the band structure with increasing number of layers ($n$=3-10) and these are displayed in Fig. 10-13.  We wish to point out that we do not include the results for multi-layers containing 7-9 layers since the trends in the band structures with increasing number of layers are similar.  We observe from Fig. 10-13 that for given number of layers ($n$), the band structures of multi-layers with AAAA, AABB and ABC stacking configurations are quite similar and on the other hand, they are different from those of multi-layer of silicene with Bernal stacking (ABAB). The reason for similarity of  band structures in AAAA, AABB and ABC is due to the similar local environment of silicon atoms in tetrahedral configurations in these stackings. However, they are different from those of multi-layer with ABAB stacking due to different arrangement of silicon atoms in this stacking. Furthermore, for a given number of layers, we observe that there exist more number of bands with characteristic dispersion of $\sigma$ bonds between Si atoms, in multi-layers of silicene with AAAA, AABB and ABC stackings as compared to those in ABAB structures.  Moreover, it can be observed that there exist two bands in multi-layers with AAAA, AABB and ABC stackings which always cross the Fermi level.  The width of these bands decreases with number of layers and hence they tend to be dispersionless as n goes beyond 6. These bands are due to the surface states since they correspond to Si atoms which are present on the surface layers. To verify the nature of these two bands, we hydrogenate the multi-layers of silicene and study the effect of hydrogenation on their band structure. The band structures of hydrogenated multi-layers of silicene with 6 layers are also plotted in Fig. 10-13 ( see bottom most right panel) for comparison with bare multi-layers. These plots clearly show the absence of bands corresponding to surface states and thus  a transition from semi-metallic to semiconductor state takes place.

Before concluding this section, we wish to make a few comments on the band structures of multi-layers with Bernal stacking.  The results presented in Fig. 13 clearly indicates that the band structures of ABAB stacked multi-layers are different from the corresponding results for other stackings. For Bernal stacking, we observe that there exist less number of bands corresponding to $\sigma$ bonds while more number of bands corresponds to the weakly bound $\pi$ bonds as compared to those of the other three stackings. These $\pi$ bands arise from Si atoms present on every layers of multi-layers of silicene. Therefore, the saturations of Si on surface layers with hydrogen atoms do not cause the system to undergo a transition from semi-metallic to semiconductor state. 

\section{Conclusion}
In this work, we perform studies on geometric and electronic properties of multi-layers of silicene with four different stacking configurations using density functional theory based calculations with GGA-PBE exchange-correlation functional. The evolution of these properties of multi-layer of silicene with increasing number of layers has been studied. 

We observe from the band structures of bi-layers of silicene that they exhibit linear and parabolic dispersions around the Fermi level, as in graphene, for AA and AB stackings respectively. However, the dispersion curves are displaced with respect to the Fermi level both along $E$ and $k$ directions in the band diagram of silicene.  The calculations on multi-layers with more than three layers show that out of four different stacking configurations considered here, namely, AAAA, AABB, ABAB and ABC, multi-layer of silicene with ABC configuration possesses minimum energy. This is in contrast to the case of  multi-layers of graphene where ABAB stacking is the lowest energy configuration. Furthermore, our results show that all the stackings with tetrahedral coordination ( AAAA, AABB and ABC ) have much higher cohesive energy than that of Bernal (ABAB) stacking. We also observe that the surface atoms on the multi-layers of silicene contribute to the bands near the Fermi level in the lower energy stacking configurations AAAA, AABB and ABC. These surface states are responsible for the semi-metallic character of the multi-layers of silicene studied here. 

From our calculations, it is clear that the major differences in the properties of bi- and multi-layer of silicene from those of graphene arise due to the presence of strong inter-layer covalent bonding between the layers of the former as opposed to the weak van der Waal's interaction which exists between the layers of graphene. 

\section{Acknowledgments}
Authors thank Dr. P. D. Gupta for encouragement and support. The support and help of Mr. P. Thander and the scientific computing group, Computer Centre, RRCAT is acknowledged.

\pagebreak

\pagebreak

\end{document}